\documentclass[twocolumn]{revtex4-1}
\usepackage{amsmath}
\usepackage{amsfonts}
\usepackage{graphicx}
\usepackage{color}
\usepackage{caption}
\usepackage{subcaption}
\usepackage{wrapfig}

\definecolor{darkgreen}{rgb}{0,0.5,0}

\newcommand{\defeq}{\mathrel{\overset{\makebox[0pt]{\mbox{\normalfont\tiny\sffamily def}}}{=}}}
\newcommand{\labelleftarrow}[1]{\mathrel{\overset{\makebox[0pt]{\mbox{\normalfont\tiny\sffamily #1}}}{\leftarrow}}}

\begin{document}
\title{Topological and geometric measurements of force chain structure}
\author{Chad Giusti}
\affiliation{Warren Center for Network and Data Science\\
University of Pennsylvania}
\email{cgiusti@seas.upenn.edu}
\author{Lia Papadopoulos}
\affiliation{Department of Physics\\University of Pennsylvania}
\author{Eli T. Owens}
\affiliation{Department of Physics\\Presbyterian College}
\author{Karen E. Daniels}
\affiliation{Department of Physics\\North Carolina State University}
\author{Danielle S. Bassett}
\affiliation{Departments of Bioengineering and Electrical \& Systems Engineering \\University of Pennsylvania}
\email{dsb@seas.upenn.edu}
\date{\today}

\clearpage
\newpage
\begin{abstract}
Developing quantitative methods for characterizing structural properties of force chains in densely packed granular media is an important step toward understanding or predicting large-scale physical properties of a packing. A promising framework in which to develop such methods is network science, which can be used to translate particle locations and force contacts to a graph in which particles are represented by nodes and forces between particles are represented by weighted edges.  Applying network-based community-detection techniques to extract force chains \cite{bassett2015extraction} opens the door to developing statistics of force chain structure, with the goal of identifying shape differences across packings, and providing a foundation on which to build predictions of bulk material properties from mesoscale network features. Here, we discuss a trio of related but fundamentally distinct measurements of mesoscale structure of force chains in arbitrary 2D packings, including a novel statistic derived using tools from algebraic topology, which together provide a tool set for the analysis of force chain architecture. We demonstrate the utility of this tool set by detecting variations in force chain architecture with pressure. Collectively, these techniques can be generalized to 3D packings, and to the assessment of continuous deformations of packings under stress or strain.
\end{abstract}

\maketitle

\section{Introduction}

Densely packed granular materials exhibit a rich internal network of physical interactions \cite{Dantu-1957-CEM,Liu-1995-FFB, Radjai1996,Mueth1998,Howell1999,Majmudar2005}, which have come to be referred to as \emph{force chains} (FIG. \ref{f:packing}). The structure of these chains is thought to play a critical role in the response of the media to perturbations including acoustic propagation \cite{Makse-1999-WEM,Hidalgo2002,Somfai2005,Owens2011} and shear \cite{Howell1999}. However, the exact physical mechanisms linking mesoscale network organization to global scale bulk properties are not well understood.  Developing measurements that effectively isolate features of force chains responsible for the bulk properties of packings that induce desirable physical behaviors is a vital first step toward designing systems that exhibit those behaviors.

However, there is not currently a widely-accepted definition of what constitutes a force chain. Here, we rely on the data-driven perspective introduced in \cite{bassett2015extraction}, which offers a fundamental mapping between measurements of granular media and mathematical objects known as weighted graphs \cite{Albert2002}. Specifically, constituent particles are represented as network nodes and the normal forces between pairs of particles in contact with one another are represented as network edges whose weight is proportional to the value of the normal force (FIG. \ref{f:packing}c). Several studies of granular media have examined properties of similarly constructed networks. For example, contact loops and other topological measures have been examined in a number of systems \cite{Smart:2008aa, Arevalo:2009aa, Arevalo:2010aa, Arevalo:2010ba, Tordesillas:2010aa, Arevalo:2013aa}. Granular force networks have also been analyzed using tools from persistent homology \cite{Kondic:2012aa, Kramar:2013aa, kramar2014, maza2014, Pugnaloni:2015, Kondic:2015} as well as with network-science based measures \cite{Walker:2010aa, Herrera2011, Bassett2012, Slotterback2012, walker-soft, bassett2015extraction}. In this study, we express the network architecture as an adjacency matrix $\mathbf{A}$ whose elements $A_{ij}$ encode the normal force between node $i$ and node $j$. Using community detection techniques \cite{Porter2009,Fortunato2010} informed by a geographically-constrained null model \cite{Bassett2013Robust}, we extract subgraphs with branch-like characteristics reminiscent of force chains. 

The ability to identify force chains in a data-driven manner unearths the more fundamental problem of identifying characteristics of force chains that drive or predict bulk properties. One simple approach lies in describing the statistics of individual interactions (force-weighted contacts) between particles. Yet, these statistics do not accurately predict signal transmission through the material, likely due to their naivety with regards to mesoscale architecture or collective dynamics \cite{Owens2011}. Alternative options include statistics that directly describe the geometry or topology of the force chains, therefore capturing mesoscale architectural properties of the material \cite{Bassett2012,bassett2015extraction}. 

In this paper, we discuss three such measures and assess their ability to (i) identify force chain structure in packings of granular particles, and (ii) detect variations in force chain architecture as a function of the applied pressure. This analysis is carried out on both laboratory packings and on simulated packings in two dimensions. We begin with the previously defined ``topophysical'' statistic known as the \emph{gap factor} \cite{bassett2015extraction}, which utilizes a blend of topological and physical information to characterize force chains. Teasing apart the complex relationship between geometry and topology in the definition of the gap factor motivates us to consider a pair of statistics which are respectively purely geometric and topological in flavor: the \emph{hull ratio}, first described in \cite{huang2015friction}, which measures the density of the packing around each chain; and a novel topological statistic of a network, the \emph{topological compactness factor} (TCF), which measures physically relevant structure in the topology of the contact network underlying the force chains. The TCF is sensitive to mesoscale features like branching in chains or compact, highly-interconnected portions of the force network, where the system's local stability can be respectively weakened or reinforced \cite{tordesillas2009modeling}. 

We find that the gap factor and hull ratio are better able to extract force chain structure at each pressure than the purely topological statistic.  On the other hand, the TCF is more sensitive to pressure than either of the other two measures. Further, its underlying mathematics can be extended naturally to quantify continuous deformations of packings in variable environmental conditions, thus forming an important tool for examining heterogeneous microstructures in particulate matter.  The three statistics together provide a broad picture of the mesoscale structure of the packing, and our findings suggest that both physical and topological information is useful to consider when studying granular systems. 

\setcounter{figure}{0}

\begin{figure}[t]
  \centering
  \includegraphics[width=0.5\textwidth]{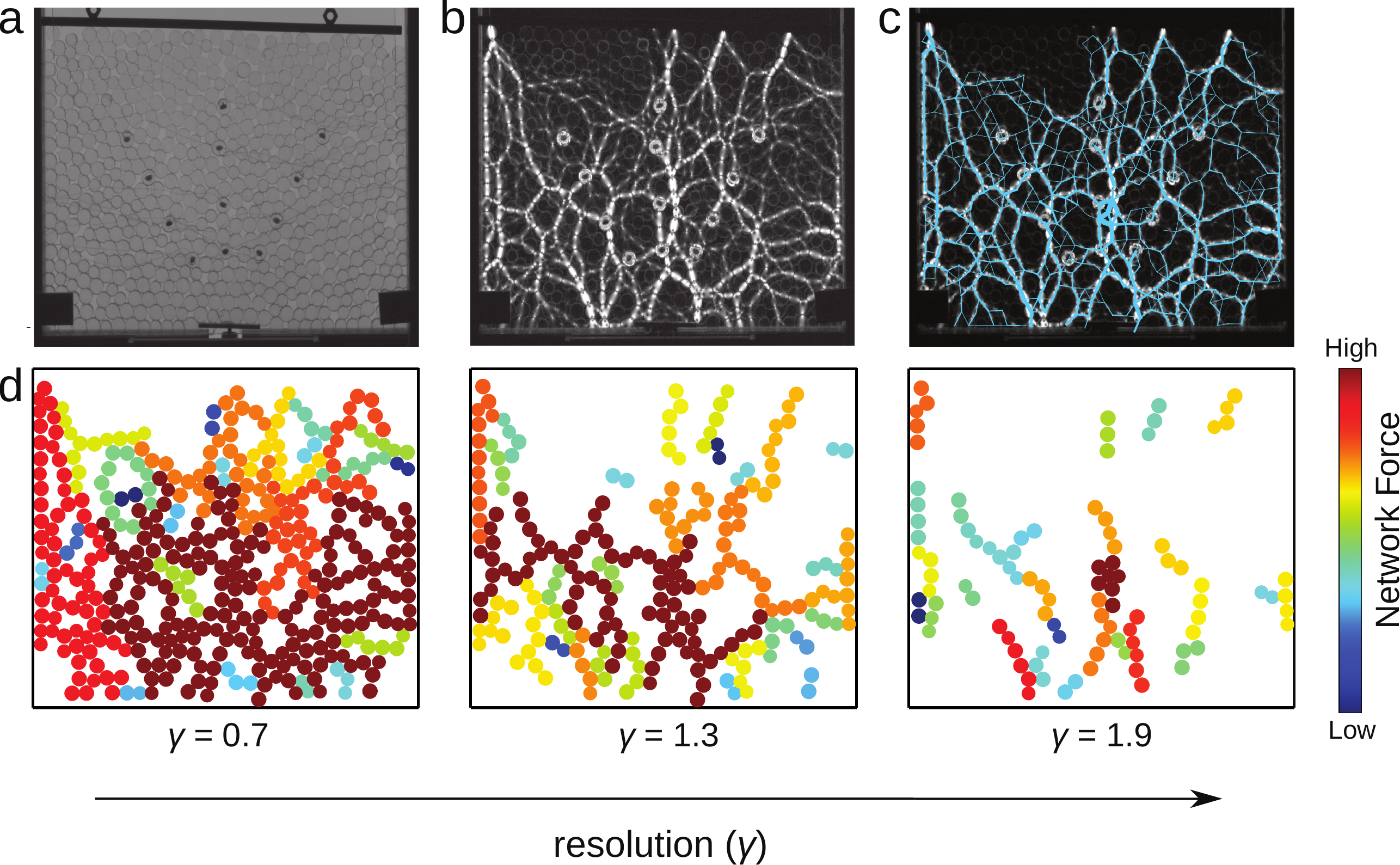}
\caption{\textbf{Community detection techniques extract force chains from granular media.} \textbf{(a)} An image of the single-layer photoelastic disk experiment. Particles are confined in two dimensions and are vertically compressed via the application of brass weights to the top of the configuration. \textbf{(b)} The internal stress pattern in the photoelastic disks exhibits force chain structure. Brightness indicates the strength of forces between particles. \textbf{(c)} The corresponding weighted graph overlaid on the photoelastic image, where the thickness of the line segments (edges) are proportional to the normal force between the two particles they connect (nodes).  \textbf{(d)} Community structure extracted from the force-weighted network as a function of the resolution parameter $\gamma$ in the modularity quality function. Colors indicate distinct communities, and warmth corresponds to {\em network force}, the amount by which the total inter-particle forces exceed $\gamma$ times the mean inter-particle force.}
\label{f:packing}
\end{figure}

\section{Methods}
\label{s:methods}

To illustrate the effectiveness of our statistics, we examine two case studies: (i) granular experiments where friction and gravity play a role, and (ii) frictionless and gravity-less simulations with periodic boundary conditions. In both cases, approximately 600 bidisperse disks with a ratio of 50:50 are confined in two dimensions and interact with each other in a Hertzian-like manner. In each packing, which is jammed under constant pressure, we measure force-contact networks for approximately the same 7 different values of confining pressure.


\subsection{Granular experiments}

We perform experiments on a vertical 2D granular system of bidisperse disks that are confined between two sheets of Plexiglas. The particles are $6.35$~mm thick, their diameters are $d_1 = 9$~mm and $d_2=11$~mm (which yields a diameter ratio of approximately 1.22), and they are cut from Vishay PSM-4 photoelastic material to provide measurements of the internal forces. These particles have an elastic modulus of $E=4$~MPa. We produce new configurations by rearranging the particles by hand, and we increase the pressure by placing additional brass weights on the top surface of the packing. The values of pressure, which we report in units of the elastic modulus $E$ (recall that the configuration is two-dimensional), are
$2.7 \times 10^{-4}E$, $4.1 \times 10^{-4}E$, $6.7 \times 10^{-4}E$, $1.1 \times 10^{-3}E$, $2.2 \times 10^{-3}E$, $3.8 \times 10^{-3}E$, and $5.9 \times 10^{-3}E$. See Refs.~\cite{Owens2011,Bassett2012,Owens2013} for additional details about the experiments.

For each of 21 particle configurations and the 7 values of pressure, we compute particle positions and forces using two high-resolution pictures of the system (FIG. \ref{f:packing}a,b).  We use one image, which we take without polarizers, to determine the particle positions and contacts. (See \cite{Puckett2013} for a description of the technique.) We take particles to be in contact if they have a measurable-magnitude force based on our photoelastic calculations. Using a second image that we take with polarizers, we then determine the particle contact forces by solving the inverse photoelastic problem \cite{Puckett2013}.


\subsection{Frictionless simulations}

We perform numerical simulations of bidisperse frictionless disks with a diameter ratio of $1.22$. Particles interact via a Hertzian potential in a box with periodic boundary conditions in both directions and zero gravity~\cite{OHernCG, OHern2003, Manning2011}. This model has been well-studied and it is significantly different from our experimental system with friction, gravity, and fixed boundaries.  We generate mechanically-stable packings via a standard conjugate-gradient method~\cite{OHernCG}. We then perform simulations for a fixed packing fraction and volume, and we analyze 20 mechanically-stable packings at each packing fraction $\phi$.  We choose the seven values of the packing fraction so that the mean pressure $\overline{p}$ at that packing fraction~\footnote {For finite systems, there isn't a bijection between pressure and packing fraction. Because the packing fraction is fixed in these simulations, there is some variance (of roughly $9 \times 10^{-5}$) in the pressure between different packings.} matches the ones in the experiments: $[\phi, \overline{p}] = [0.8499, 3 \times 10^{-4} E]$,
$[0.8521, 4 \times 10^{-4} E]$,
$[0.8560, 7 \times 10^{-4} E]$,
$[0.8621, 11 \times 10^{-4} E]$,
$[0.8760, 22 \times 10^{-4} E]$,
$[0.8927, 38 \times 10^{-4 }E]$, and
$[0.9106, 59 \times 10^{-4} E]$,
where the modulus $E$ is defined as the energy scale for the Hertzian interaction $\epsilon$ divided by the mean Vorono\"i area of a particle in the packing. The lowest value of $\phi$ provides a data point for a jammed packing that is less dense than what is accessible in our experiments.

\section{Force chain extraction via community detection}
\label{s:forcechains}

Following \cite{bassett2015extraction}, we represent the forces between particles using a weighted graph and we apply community detection techniques to extract putative force chains from the resulting network. First, we construct a simple, unweighted \emph{contact graph} $\mathbf{B}$ which has as nodes the particles in the system and which has an edge between nodes precisely when the corresponding particles are in contact. We extend this to a weighted \emph{force graph} $\mathbf{W}$ by including edge weights $W_{ij}$ equal to the normal force between the particle $i$ and $j$.

To extract force chains from the force network $\mathbf{W}$, we need to find sets of particles with strong interparticle forces among themselves. We accomplish this by using community detection techniques \cite{Porter2009,Fortunato2010} to partition the vertices into groups (or \emph{communities}) in a fashion that maximizes a modularity quality function \cite{Newman2006}. The particular function we utilize is given by
\begin{equation}
Q = \sum_{i,j} [W_{ij} - \gamma P_{ij}]\delta(g_i, g_j)\\,
\end{equation}
where node $i$ is assigned to community $g_i$, node $j$ is assigned to community $g_j$, $\delta$ is the Kronecker delta function, and $P_{ij}$ is a null model. The parameter $\gamma$ is the structural \emph{resolution} parameter that can be used to tune the number of communities detected: small values of $\gamma$ will produce fewer larger communities, while larger values of $\gamma$ will produce many smaller communities. We optimize modularity for several values of the resolution parameter between $\gamma = 0.1$ and $\gamma = 2.1$ in steps of 0.2. As in \cite{Bassett2013Robust,bassett2015extraction}, we employ a \emph{geographical null model} given by
\begin{equation}
P_{ij} = \rho B_{ij}\\,
\end{equation}
where $\rho$ is the mean edge weight in the network, to reflect the constraints on network structure induced by the two-dimensional packing. Maximizing the modularity quality function partitions the particles in the system into network communities, and by tuning the resolution parameter appropriately, we can identify groups of particles that are visually reminiscent of force chains (FIG. \ref{f:packing}d). As the modularity function is non-convex, we use a locally greedy (``Louvain''-like \cite{Blondel2008}) algorithm \cite{genlouvain} to obtain approximations of the optimal partition of nodes into communities. We report results across twenty such optimizations of the modularity quality function per packing for the experimental system, and across one-hundred optimizations per packing for the simulations. 

\section{Characterizing force chains}
\label{s:statistics}

As we vary the resolution parameter $\gamma$, the character of the force chains we extract from a packing change dramatically (FIG. \ref{f:packing}d). In order to provide consistent characterizations of structure, we wish to select an ``optimal'' resolution at which the results of the community detection strongly resemble our heuristic notion of what a force chain should be: heterogeneous and branch-like. Defining statistics to quantify such a branching architecture is an open area of investigation \cite{bassett2015extraction}. In what follows, we define three different statistics that we use to characterize the sets of particles found from community detection. We then discuss the ability of these statistics to identify an optimal resolution parameter for force chain extraction, and consider their sensitivity to the pressure applied on the system. 

\subsection{The gap factor: a previously defined ``topophysical'' statistic}
\label{s:gapfactor}

\begin{figure}
  \centering
  \includegraphics[width = 0.45\textwidth]{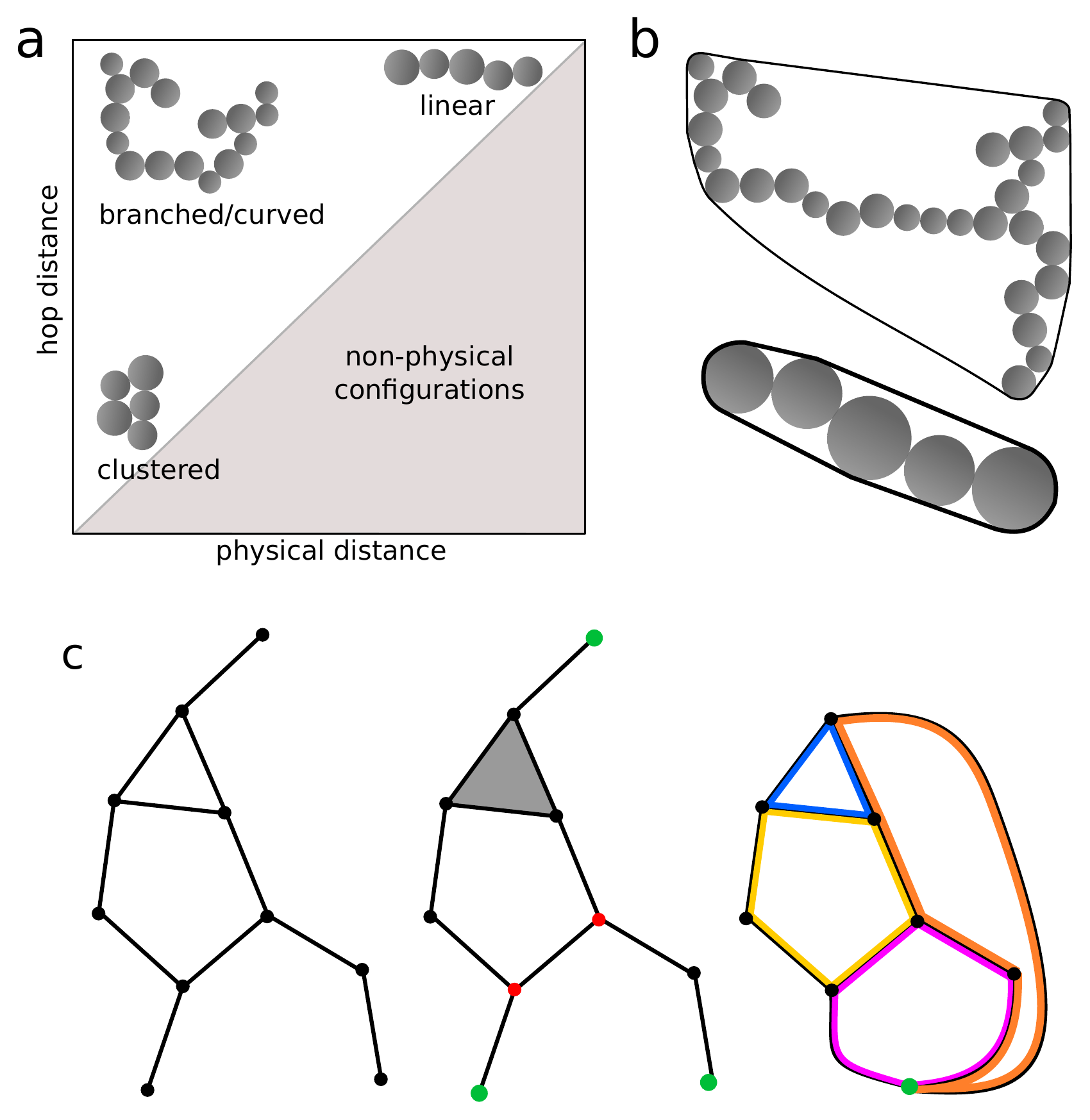}
\caption{\label{f:measures} \textbf{Three metrics of force chain structure.} \textbf{(a)}  The \emph{gap factor} of a force chain measures the discrepancy between the topology of the contact network of a force chain and its physical layout. Force chains for which the two are poorly correlated display branching or curving behavior. \textbf{(b)} The {\em hull ratio} of a force chain is the ratio of the area of the constituent particles to those of its convex hull. Curved or branched chains (top) have low hull ratio, while straight or clustered chains (bottom) have a high hull ratio. \textbf{(c)} The {\em topological compactness factor} (TCF) measures mesoscale connectivity properties of the contact graph that contain physically relevant information. (left) Binary contact graph for a force chain. (middle) A {\em compact} region (grey triangle) is a cluster of particles in the chain which are connected to all of their neighbors in a triangular grid, while {\em branch points} (red) are particles with more than two neighbors, none of which share an edge. (right) After collapsing the {\em leaves} (green) in the contact graph to a single vertex, we can enumerate {\em fundamental cycles} (purple, yellow, blue, orange) in the graph to measure the prevalence of compact regions versus branch points of the chain. }
\label{f:measures}
\end{figure}

Before discussing new statistics, we recall the gap-factor, which measures the correlation between physical distance (measured using the standard Euclidean metric) and ``hop distance'' (which counts distance measured only along network edges). In this way, the gap factor combines the physical and topological information about the system into a ``topophysical'' estimate of the branch-like structure of the force chains. Force chains with gaps, branches, and rings have a larger hop distance than physical distance, and the presence of such complicated shapes decreases the correlation between the physical and hop distance (FIG. \ref{f:measures}a).

The physical distance between nodes in a force chain is the usual Euclidean distance matrix, which we denote $\mathbf{L}_p$. The \emph{hop distance matrix} $\mathbf{L}_t$ of a community is obtained by restricting the binary contact network $\mathbf{B}$ to nodes assigned to the community $c$ to obtain a submatrix $\mathbf{B}^c$. The $(i, j)$-entry of $\mathbf{L}_t$ is then given by the minimum number of edges traversed in any path through $\mathbf{B}^{c}$ from node $i$ to node $j$. 

To obtain a global statistic, we weight each community by its size so that larger communities are weighted more heavily than small communities to reflect their relative influence on the system's behavior. We define the \emph{gap factor} $g_c$ of a community $c$ to be
\begin{equation}
	g_c = 1-\frac {r_c \, s_c}{s_{\mathrm{max}}}\,,
\label{eq:gapfactor}
\end{equation}
where $r_c$ is the value of the Pearson correlation coefficient between the strictly upper triangular entries of $\mathbf{L_{t}}$ and those of $\mathbf{L_{p}}$, and $s_{\mathrm{max}}$ is the size of the largest community. We define the systemic gap factor as
\begin{equation}
	g_s = 1 - \frac{1}{n_{>1}} \sum_{c} \frac {r_c \, s_c}{s_{\mathrm{max}}}\,,
\label{eq:gapfactorsys}
\end{equation}
where the quantity $n_{>1}$ is the total number of communities, excluding singletons.

\subsection{The hull ratio}

\begin{figure*}
  \centering
  \includegraphics[width=2\columnwidth]{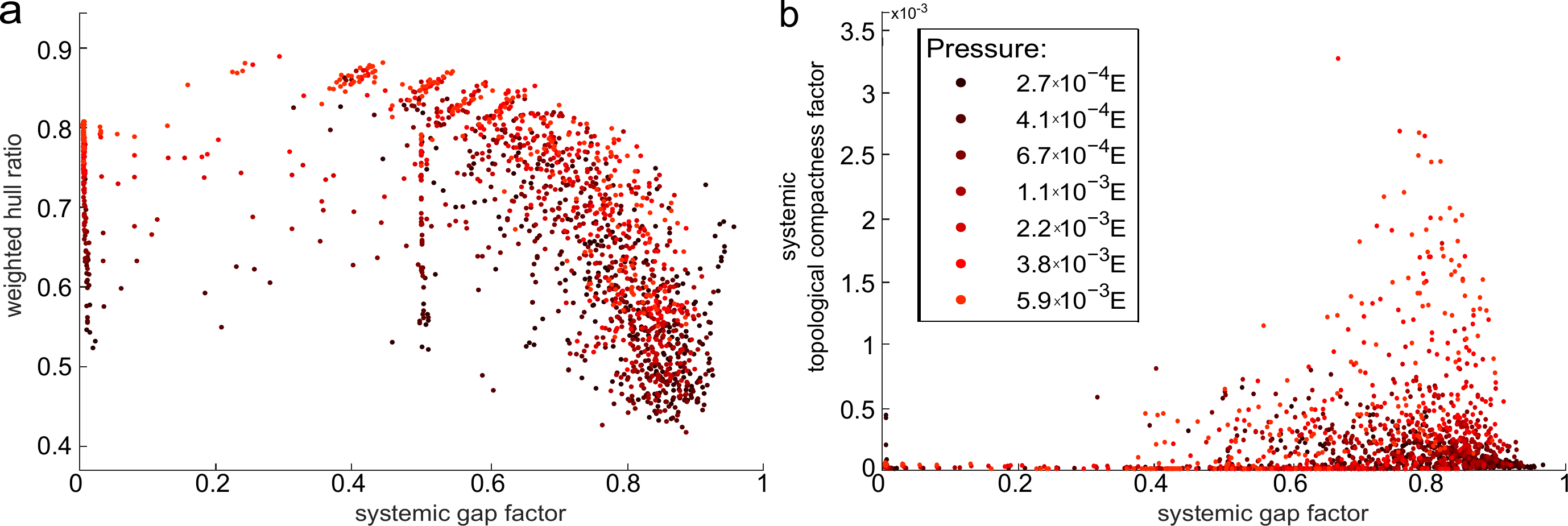}
  \caption{\textbf{Comparisons of hull ratio and TCF to gap factor.} Scatter plots of \textbf{(a)} hull ratio and \textbf{(b)} TCF against the gap factor. Points correspond to the mean of each measure across optimizations of the modularity quality function for each packing, each pressure, and choice of the resolution parameter between 0.1 and 2.1. The hull ratio is strongly correlated with the gap factor (Spearman's $\rho = -0.72$), while the TCF is more weakly correlated with the gap factor ($\rho = -0.37$). The striped regions in (a) correspond to packings with small numbers of communities, and are explained in detail in the text. }
  \label{f:measure}
\end{figure*}

To complement the topophysical gap factor, it is useful to consider a purely geometric quantity that depends only on the relative locations of particles in physical space, and therefore does not use notions of network topology.  A simple measure of how tightly packed a group of particles is obtained by taking the ratio of the total area of the particles to the area of the convex hull of the particles (FIG. \ref{f:measures}b). Following \cite{huang2015friction}, we define the \emph{hull ratio} of a community $c$ to be 
\begin{equation}
R(c) = \frac{\text{Area}_\text{particles}(c)}{\text{Area}_\text{convex hull}(c)}.
\end{equation}
With this definition, $R(c)$ will be close to unity for more compact and linear structures, whereas groups of particles with heterogeneous force chain geometry will have smaller values of $R(c)$. To obtain a summary statistic for the structure of a packing as a whole, we define the \emph{weighted hull ratio}
\begin{equation}
R_s = \frac{\sum_{c} R(c) \cdot s_c}{\sum_c s_c}
\end{equation}
given by the average of the hull ratios for all the communities, weighted by the number of particles in the community (again excluding singletons).


\subsection{The topological compactness factor}
\label{s:topology}

To complement the previously defined gap factor (a ``topophysical'' statistic) and hull ratio (a physical statistic), we turn now to defining a purely topological statistic that is sensitive to the tendency of the force chain to contain compact regions or to branch (FIG. \ref{f:measures}c). 

To motivate the development of this new statistic, we first note that the contact structure of chains is strongly reflective of the organization of forces within the system, and that the mesoscale properties of the packing near these compact regions or branch points is of particular structural significance. A {\em compact} cluster of particles is one in which the constituent particles exert pairwise forces on one another, forming a roughly triangluar tiling in the 2D system which necessarily constrains its motion. A {\em branch point} occurs when the chain diverges into two or more locally distinct paths across a single particle (FIG. \ref{f:measures}c). There is theoretical work suggesting that compact regions in the force network increase rotational stability of the system \cite{rivier2006extended}, and we posit that branch points increase the propensity for the chain structure to slip under perturbation of the system. Here, we propose a statistic to explicitly assess these two tendencies: the \emph{topological compactness factor} (TCF). 

The TCF is conceptually related to the graph-theoretic clustering coefficient, which measures the percentage of neighbors of each vertex that share an edge, however the TCF focuses attention on the mesoscale structure of each chain rather than its behavior at particular vertices. The TCF derives from techniques drawn from algebraic topology, and is in fact a summary statistic of a much more powerful measurement of the structure of the chain in terms of the {\em homology} of its {\em clique complex}. However, as it is possible to describe the numerical TCF in purely graph-theoretic terms in the case of a static 2D packing, such as those considered in this paper, we instead provide such a description here and leave the more complete definition to the Appendix \ref{s:algtop}.

While compact regions can be relatively easily extracted from the contact graph by the triangular connectivity patterns they induce, counting branches is an inherently non-local process. Some branches are most easily identified by counting the number of leaves of the force chain graph (FIG \ref{f:measures}c (middle)), but other branches may combine to form loops which are entirely internal to the chain. In order to put branch points into the same framework as compact regions, we begin by modifying the graph so that both can be expressed in the language of ``cycles''. Specifically, let $c$ be a community in the packing, thought of as its underlying contact graph $\mathbf{B}^c$. Recall that a {\em leaf} in $\mathbf{B}^c$ is any vertex that shares an edge with precisely one other vertex (FIG. \ref{f:measures}c (middle, green)). Construct a new graph $\widetilde{\mathbf{B}}^c$ from $\mathbf{B}^c$ in two steps: first, introduce a new vertex $\pi$ and add an edge from vertex $i$ to $\pi$ precisely when there is some leaf $\ell$ of $\mathbf{B}^c$ whose unique edge is to vertex $i$; second, delete all leaves from $\widetilde{\mathbf{B}}^c$ (FIG. \ref{f:measures}c (right)).
 
Recall that a {\em cycle} in a graph G is a path that begins and ends at the same vertex (FIG. \ref{f:measures}c (right)). A common difficulty when enumerating cycles in graphs lies in determining which cycles to count, as illustrated by the common ``How many rectangles do you see?'' puzzle. Here, we wish only to consider a {\em basis} for the set of cycles under the operation of {\em composition}, and so use the matroid-theoretic notion of {\em fundamental} cycles. Recall that a \emph{forest} is a graph without cycles and that a {\em spanning} forest for $G$ is a forest obtained by removing edges from $G$ until no cycles remain, without changing the number of connected components. The number of fundamental cycles in $G$, written $F(G)$, is defined to be the number of edges which must be removed from $G$ to create a spanning forest\footnote{This quantity is well-defined with regard to the choice of spanning forest: adding an edge to a spanning forest necessarily introduces a cycle and thus forces the removal of a different edge, while removing an edge increases the number of connected components, necessitating the reintroduction of another edge.}. 

Observe that cycles of length 3, also known as 3-cliques, are present precisely where three particles in the packing are all in pairwise contact (FIG. \ref{f:measures}c (right, blue)). Such a structure will be our indication that the cycle is induced by membership in a compact region, rather than a branch point; denote by $T(G)$ the number of such cycles in $G$. 

Finally, we define the {\em topological compactness factor} of a force chain $c$ to be
\begin{equation}
TCF(c) = \frac{T(\widetilde{\mathbf{B}}^c)}{F(\widetilde{\mathbf{B}}^c)}.
\end{equation}
and set the TCF to be zero when $F(\widetilde{\mathbf{B}}^c) = 0$. Chains for which the TCF is large will tend to consist of compact, highly interconnected regions of particles, while those for which the TCF is small will tend to be long, branching chains. In order to compare packings, we define a summary statistic (following \cite{bassett2015extraction}), which we refer to as the \emph{systemic topological compactness factor}
\begin{equation}
T_{s} = \frac{1}{n} \sum_{c}TCF(c) \cdot \frac{s_c}{s_{\text{max}}}
\end{equation}
where $n$ is the number of communities $\{c\}$, $s_c$ is the size of community $c$ and $s_{\text{max}}$ is the size of the largest community.

\subsection{Comparison of community measures}

It is interesting to compare the previously defined gap factor to the geometric hull ratio and topological compactness factor. In FIG. \ref{f:measure}, we show scatter plots of (a) the hull ratio vs. the gap factor and (b) the compactness factor vs. the gap factor. In both plots, points correspond to the mean of each measure across optimizations of the modularity quality function for each packing, pressure, and choice of the resolution parameter between 0.1 and 2.1. Though the hull ratio and gap factor do not appear linearly correlated, it is evident from panel (a) that there is a relationship between their values and we find that they are indeed strongly Spearman correlated ($\rho = -0.72$).  On the other hand, the TCF is only weakly correlated with the gap factor ($\rho = 0.37$), and this topological statistic thus likely captures different information about community structure than either the gap factor or the hull ratio measures.

The scatter plot of the hull ratio against the gap factor contains two additional features that stand out by eye: the vertical bands at $g_{s} \approx 0$ and $g_{s} \approx 0.5$ and the diagonal striations near the top of the figure. Investigation of the vertical stripes shows that they occur at small values of the resolution parameter. At low $\gamma$, the hull ratio is sensitive to the packing pressure, which can be seen by the vertical stratification of colors in both bands. Moreover, the spread of the vertical stripes along the horizontal axis can be understood from the definition of the weighted gap factor in EQN. \ref{eq:gapfactorsys}. At low resolutions, there will either be a single large, compact community or one large, compact community with a few very small communities dispersed throughout the packing. The exact number of communities detected at a given pressure and resolution parameter will in part be dependent on the structure of the particle configuration. If there is a single large community that contains the majority of the particles, then $r_{c} \approx 1$ and the gap factor will be close to zero (the first vertical band).  However, the gap factor is highly sensitive to the number and size of communities; if a given configuration contains one large community and a just a couple of small communities of size two or three, the weighted gap factor will be noticeably affected. The correlations $r_{c}$ will still be close to one for all communities, but the division by the number of communities (which in this situation is $n > 1$) will shift the gap factor values to the right (this is the second vertical band). For example if $n = 2$ and $r_{c} \approx 1$ for all communities, then $g_{s}$ will be close to $0.5$ when one community is large (size $s_{max}$) and the other community satisfies $s_{c} \ll s_{max}$.

A similar issue causes the diagonal striations seen near the top of FIG. \ref{f:measure}a. Variations in packing structure will again cause the number and size of communities to differ slightly between different experimental configurations at the same pressure. When the number and size of communities is few (i.e. at high resolutions), the gap factor is sensitive to such small differences and its values can take on a large spread (this causes the set of diagonal bands across the gap factor axis). The hull ratio is more robust against these small differences, and so for fixed resolution parameter and pressure, its values are contained to a smaller range.  Both of these features disappear at intermediate values of $\gamma$ when there are more communities and many communities are intermediately sized. An alternative definition of community averaging in the weighted gap factor could alleviate the large spread in values that can occur for similar packings at low and high resolutions.

\section{Force chain identification and sensitivity to pressure}
\label{s:force_chains&pressure}

\subsection{Force chain identification}
\label{s:force_chains}

We now apply each of these statistics to collections of laboratory and simulated packings under a variety of pressure conditions ranging from $2.7\times 10^{−4}E$ to $5.9\times 10^{−3}E$. Our first goal is to examine how well these measures can identify force chain architecture from the results of the community detection methods. We compute the value of each statistic over the eleven different resolution parameters between $\gamma = 0.1$ and $\gamma = 2.1$, and report averages over all optimizations and packings. Beginning with the gap factor, we follow the method described in \cite{bassett2015extraction} and extract force chains at each pressure by identifying an optimal value of $\gamma$ that maximizes the gap factor (FIG. \ref{f:experimental_simulated}a). For the experimental data set, we find that the optimal region occurs between $\gamma =  0.7$ and $\gamma = 0.9$, and these resolutions also give rise to communities that strongly resemble the force chain architecture seen in the photo elastic disk experiments (FIG. \ref{f:packing_structure}a). One can identify optimal resolutions for the simulations as well (FIG. \ref{f:experimental_simulated}b), and we find that these values occur predominantly between $\gamma = 1.1$ and $\gamma = 1.3$. 

A similar process can be carried out with the hull ratio, but this time we take the optimal value of $\gamma$ at each pressure to be that which minimizes ${R_s}$. The resolutions that minimize the hull ratio should give rise to more heterogenous structure and less compact and linear structure. The optimal values lie between $\gamma = 0.7$ and $\gamma = 0.9$ for the experimental data (FIG. \ref{f:experimental_simulated}c), and between $\gamma = 0.9$ and $\gamma = 1.1$ for the simulations (FIG. \ref{f:experimental_simulated}d). The extrema of the hull ratio are very well defined for the laboratory packings (even more so than those of the gap factor), and it is thus clear where the optimal resolution parameters lie. In addition, the force chain region identified from the hull ratio agrees well with that identified from the gap factor. These findings suggest that physical distance information alone -- as instantiated in the hull ratio -- may be sufficient for extracting force chain structure once we have constructed communities using the inter-particle contact forces.

The TCF depends only on the network topology and does not take into consideration any physical distances. It is somewhat less straightforward to use this quantity to identify the force chain region of the structural resolution parameter $\gamma$ than it is with either the gap factor or hull ratio. In the experimental data, the resolution parameters that correspond to maxima in the TCF occur at values similar to those found to maximize the gap factor and minimize the hull ratio (FIG. \ref{f:experimental_simulated}e). In contrast, in the simulated data the resolution parameters that correspond to maxima in the TCF occur at very small resolutions, but local maxima correspond to values found to maximize the gap factor and minimize the hull ratio (FIG. \ref{f:experimental_simulated}f).

Our analysis shows that from the three statistics we examined, those that include notions of physical shape are most adept at quantifying the presence of branching and other heterogeneities known to occur in force chain structure. Both the gap factor (which uses a combination of physical and topological distance information) and the hull ratio (a purely geometric measure) have well defined extrema at resolution parameters that result in groups of particles reminiscent of the force chains we see by eye.  On the other hand, the purely topological TCF is less able to identify these optimal resolutions in the experimental data. Based on the outcome of our examination, we consider the force chain region for the laboratory data to occur between resolution parameters of $ \gamma = 0.7 $ and $ \gamma = 0.9$. For the simulated data, we take a larger range for the force chain region, between $ \gamma = 0.9 $ and $ \gamma = 1.3 $.

\subsection{Sensitivity to packing pressure}
\label{s:detect_pressure}

\begin{figure*}

  \includegraphics[width=1\textwidth]{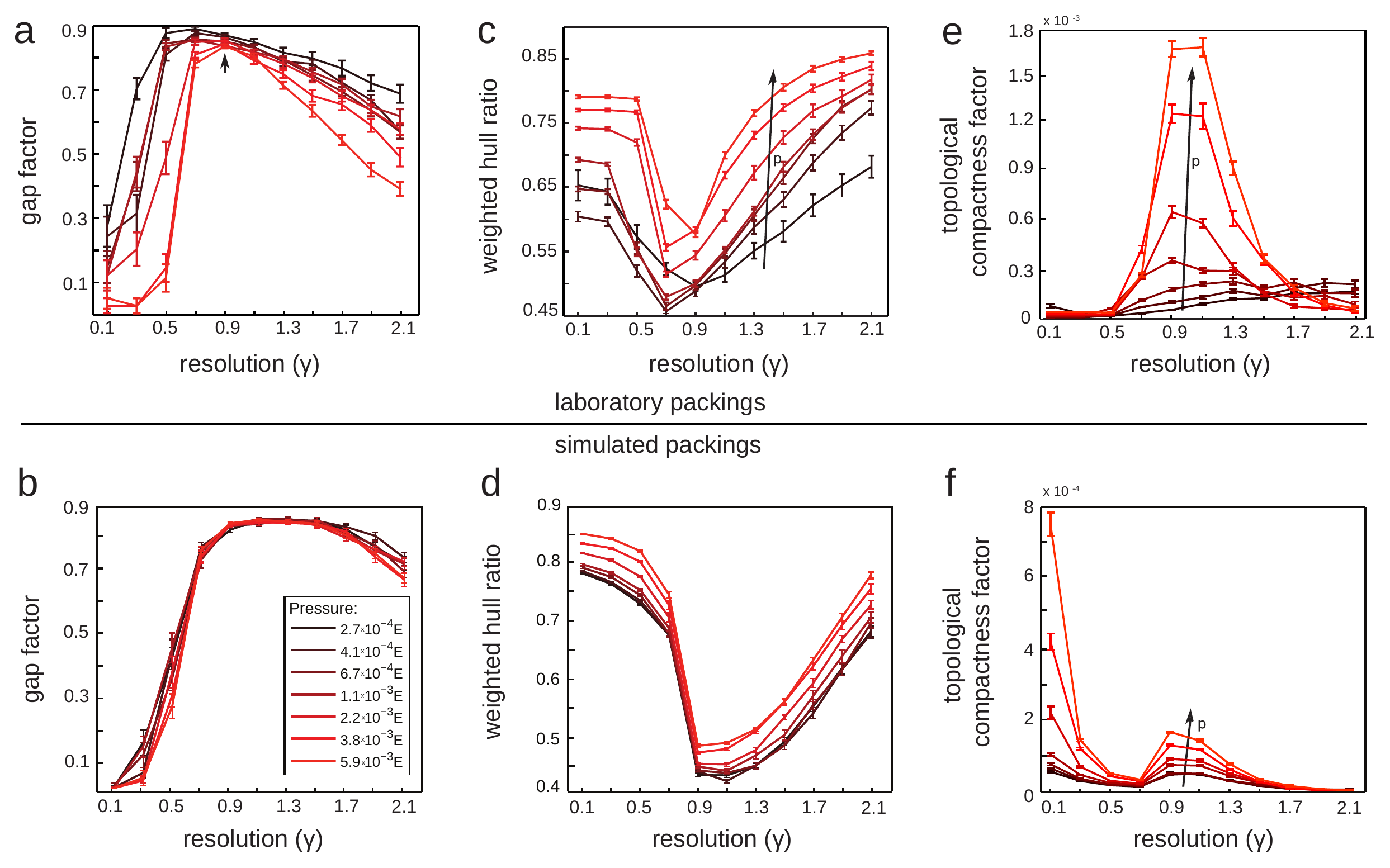}

  \caption{\textbf{Measurements of force chain structure across resolutions.} \textbf{(a, b)} The systemic gap factor as a function of resolution parameter $\gamma$ across varying pressure on laboratory \textbf{(a)} and simulated \textbf{(b)} packings. Curves are means across trials, and error bars indicate one standard deviation. The maxima of the gap factor correspond to the optimal resolution parameters for identifying community structure reminscent of force chains; but in both the laboratory and simulated packings, the gap factor is relatively homogeneous across pressure. \textbf{(c, d)} The weighted hull ratio as a function of resolution parameter $\gamma$ and pressure for the laboratory \textbf{(c)} and simulated \textbf{(d)} packings.  The optimal $\gamma$ is that which minimizes $R_s$, and agrees relatively well with the optimal $\gamma$ found from the gap factor in both experiments and simulations. However, the laboratory weighted hull ratio can distinguish pressure across packings for large values of $\gamma > 1.3$. The simulated hull ratio also partially distinguishes pressures for high values of $\gamma$, but is less well differentiated.  \textbf{(e, f)} The systemic topological compactness factor $T_s$ as a function of the resolution parameter $\gamma$ for the laboratory \textbf{(e)} and simulated \textbf{(f)} packings. $T_s$ provides a strong discrimination of packing pressures at the optimal resolution obtained from the hull ratio. The high values of the TCF for $\gamma \approx 0.1$ is the result of the small number of compact communities which rapidly fragment as $\gamma$ increases.}
  \label{f:experimental_simulated}

\end{figure*}

\begin{figure*}

  \includegraphics[width=0.75\textwidth]{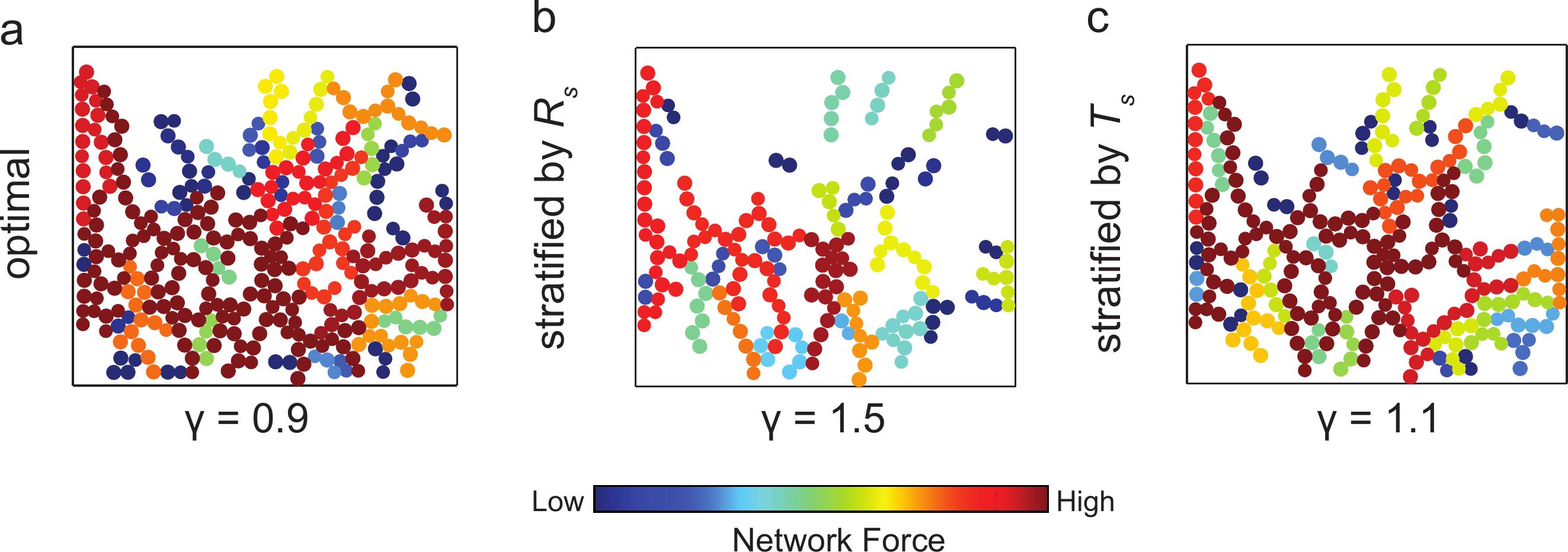}

  \caption{\textbf{Community structure for the experimental packings at the optimal and pressure-stratifying resolution parameters.} \textbf{(a)} Near the optimal resolution parameter, found by maximizing the gap factor or minimizing the hull ratio, the communities have force chain structure. \textbf{(a)} For larger values of $\gamma$, communities are small and sparsely distributed. However, the hull ratio indicates that their structure is reflective of packing pressure. \textbf{(c)} The topological compactnes factor $T_s$ best distinguishes packings by pressure at $\gamma \approx 1$, where the communities are long and branch irregularly and are also reminscient of force chains.} 
  \label{f:packing_structure}

\end{figure*}

In the previous section, we explored the ability of three statistics to identify force chain-like structure in experimental and simulated granular packings over a range of pressures. Another important question is if those same measures can detect differences in packing architecture as a function of the confining pressure, or if they are insensitive to pressure changes. As observed in \cite{bassett2015extraction}, under increasing pressure, the communities in the force network have (on average) smaller geometric radius for a given resolution parameter $\gamma$. In the resolution range where the resulting communities qualitatively resemble force chains (FIG. \ref{f:packing_structure}a), it is natural to expect features of the packings that rely on force chain structure to be visible.  

We observe that $R_s$ and $T_s$ both stratify real and simulated packings across pressures, though they do so in different ranges of the resolution parameter. In contrast, the gap factor does not distinguish pressures particularly well at any resolution (FIG. \ref{f:experimental_simulated}a, b), as observed in \cite{bassett2015extraction}. For the simulated packings especially, the curves corresponding to each pressure collapse onto one another, making it difficult to identify the relative pressure on the system from the measured statistics. For the laboratory packings, the hull ratio distinguishes pressure best when $\gamma > 1.3$ (FIG. \ref{f:experimental_simulated}c). In particular, we observe that in the stratifying regions, the hull ratio increases as pressure increases. This behavior agrees with the notion that increasing pressure generally leads to more compact structure. However, it is important to point out that the stratifying range lies outside the resolution range where communities are most representative of the force chains we observe visually.  In FIG. \ref{f:packing_structure}b, we show an example of the community structure in the region where the hull ratio can distinguish pressure. The modules are smaller in size and number compared to those extracted at the optimal resolution (FIG. \ref{f:packing_structure}a). On the other hand, the purely topological statistic best distinguishes pressure near $\gamma \approx 1$ (FIG. \ref{f:experimental_simulated}e), which is much closer to the force chain region identified in the previous section. In this $\gamma$ range, the community structure in the experimental packings resembles the types of force chains one might pick out by eye (FIG. \ref{f:packing_structure}c).  For the simulated packings, the hull ratio can distinguish between most pressures near the minima (i.e. the force chain region), but the relative separation of the curves is quite small (FIG. \ref{f:experimental_simulated}d). Of the three statistics, the TCF is the most sensitive to pressure differences (especially to higher pressures) near the force chain region in the simulations (FIG. \ref{f:experimental_simulated}f).

To understand these results, observe that in both the laboratory and simulated packings, the number of distinct force chains detected drops, while their size increases with pressure (\cite{bassett2015extraction}, FIG. 6). In the resolution range where communities are most representative of force chains, at higher pressures the resulting chains are long, but branch irregularly. The gap factor, defined as the correlation between the hop distance and the geometric distance between nodes in each force chain, will have difficulty detecting the difference between long chains with several short branches and those that are essentially linear. However, the topological measurement treats all branches equally and, as a result, can detect the increase in such motifs. Once $\gamma > 1.3$, under high pressure the tendency of strong communities will be to form tightly packed clumps, as opposed to curved structures with large convex hulls. In the experimental setting, the friction between particles can effectively protect branching chains that would otherwise collapse in frictionless numerical settings.

\section{Discussion}
\label{s:conclusions}

A necessary first step toward understanding the physical properties of packed granular materials is the development of statistics that provide insight into the structure of models. Here, we have extended the data driven, network theoretic approach to the study of force chains initiated in \cite{bassett2015extraction}, extracting putative chains using techniques from community detection. In order to understand their structure, we consider three statistics: the previously defined \emph{gap factor}, along with a purely geometric measure called the \emph{hull ratio} which captures much of the same information without relying on network topology, and a novel algebraic-topological measurement called the \emph{topological compactness factor} which distinguishes pressure in both experimental and simulated packings, showing that the purely topological properties of the force chains vary in a predictable manner with changes in pressure.

\paragraph{Network-Based Tools for Assessment of Material Architecture} 

Network science provides a natural framework in which to both represent and characterize granular materials. In this paradigm, particles are nodes in a graph and inter-particle forces are edges. Once such a graph is constructed, a number of network-based measures can be used to assess the material architecture at different spatial scales. Network science approaches seem to be especially advantageous in their ability to probe mesoscale features of granular media, of which force chain structure is a prime example. This heterogenous architecture constrains the mechanical properties of granular media under compression and shear \cite{Howell1999}, and it may also constrain the heterogeneous and nonlinear nature of acoustic signal transmission that particulate and continuum models often fail to describe \cite{Digby:1981, Goddard:1990, Liu:1994, Jia:1999,Maske:1999,Velicky:2002,Hidalgo2002,Maske:2004, Somafi:2005,Owens:2011}. 

A primary aim of the present work was to thus suggest three distinct measures for the characterization of crucial mesoscale organization in granular networks, and explore their sensitivity to pressure differences in experimental and laboratory systems. A number of previous studies have investigated intermediate scale features of granular systems as well. A large focus has been on the role of contact loops of three or more particles under various conditions. For example, using a graph representation of a granular packing, \textcite{rivier2006extended} demonstrated that the stability of the packing was related to the dynamical frustration of odd cycles in the network. In a study on the co-evolution of cycles and force chains \cite{Tordesillas:2010aa}, 3-cycles were found to be stabilizing mesoscale structures in the lead-up to force chain buckling \cite{Tordesillas:2007aa}. Cycles may also characterize the jamming transition; in particular the number of triangles grows suddenly at the critical packing fraction in simulations of isotropically compressed granular systems, suggesting that the presence of 3-cycles is characteristic of a stable, rigid state \cite{Arevalo:2010aa}. The evolving structure of contact loops has been investigated in simulations of tilted granular packings as well \cite{Smart:2008aa}.  In a study on simulations of tapped granular packings, \textcite{Arevalo:2013aa} found that polygons in the network could distinguish between equilibrium states of the same packing fraction. Spatially embedded communities probe another mesoscale feature of granular packings. Community structure can predict certain features of acoustic transmission in experimental systems \cite{Bassett2012}, and when determined with a physically informed null model, network communities result in force chain-like structure which can be subsequently analyzed and compared across varying pressures and different types of systems \cite{bassett2015extraction}.  Examples of other recent studies on granular media that have utilized a network-based analysis to probe local, intermediate and global scales include \cite{Arevalo:2009aa, Walker:2010aa, Herrera2011, Slotterback2012, Walker:2012aa, Lopez2013, walker-soft, Navakas2014}. 

The findings mentioned above demonstrate that network science is a useful tool in understanding the complex material properties of granular media. In particular, they emphasize the general need to define new measures that can quantify heterogeneous architectures between the particle-level and bulk scales. An exciting direction for future work would be to examine how the physically informed method of community detection introduced in \cite{bassett2015extraction} and used in this study, as well as the statistics introduced here and in prior work, can explain or predict interesting features of acoustic propagation in granular experiments. 

\paragraph{New Insights from Algebraic Topology}

Algebraic-topological methods have previously been applied to the study of force networks in granular media \cite{kramar2014}. The previous approach involved application of \emph{persistent homology} to the complete, weighted force network of the system (as well as related networks, such as weighting by tangent forces) to extract quantitative measures of the global network structure. Studying these global statistics, the authors see significant differences in the evolution of components and loops in the network as packing density changes, and observe a severe decrease in the rate of change of these measures as the system passes the jamming transition. Here, instead, we describe a highly local measure of rigidity in structure of individual force chains, and then aggregate these to obtain an estimate of rigidity and stability throughout the entire packing. 

This algebraic-topological approach provides a powerful tool set that that can be extended in a variety of ways. Using the definition of the statistic in terms of the general framework of cliques and chain complexes in Appendix \ref{s:algtop}, it is straightforward to extend this approach to 3D packings by including cliques on four vertices and computing $H_2^{X, \ell}(\mathbf{B})$ and its corresponding normalization. Even more strikingly, however, because we have lifted the computation to the level of vector spaces with bases defined in terms of elements of the system, we can study how these vector spaces evolve as the system does by translating these changes into linear maps. Techniques from topology such as \emph{zig-zag persistence} \cite{carlsson2010zigzag}, which describes how to ``glue together'' homology from related networks, can then be used to study the evolution of homological features, thus providing deeper insights into the properties of the dynamic system than simple numerics. 

\paragraph{Implications for Network Design} 

In future studies, we expect network-based approaches to directly impact the field of material design, which aims to engineer novel materials with desired and optimized physical properties. One notable advantage of network science is that the mathematical tools are agnostic to the physical makeup of the system under inspection, thus offering a framework in which to study a diversity of materials ranging from granular media to biological tissues \cite{Zhang:2015}. This versatility may be especially promising for an emerging branch of material design concerning so-called meta-materials \cite{Liu:2011, Turpin:2014, Lee2012}, which are developed via precise control of shape, geometry, orientation, and arrangement, rather than by choice of specific material composition. Meta-materials come in many forms; for example, some are designed with tailored mechanical and acoustic properties \cite{Lee2012, Rocklin:2015, Fang:2006, Nicolaou:2012aa}, others with desired electromagnetic properties \cite{Liu:2011, Simovski2012, Smith:2004aa}, and more. Graph theoretic approaches have potential to aid in the development of numerous materials that have underlying network topology.  

From a material design perspective, a second advantage of network science is that it provides a means to quantify structure on all scales, including mesoscale architecture in heterogeneous and disordered materials. As noted previously, this type of structure is often crucial for determining material properties. Furthermore, it is not limited to the granular regime; for example, heterogeneities and structure on the intermediate scale are also known to be important in biological materials (for some examples, see \cite{Pong:2011, Quinn:2011, Han:2013, Sporns:2013aa, Sporns:2014aa, Zhao:2014, Zhang:2015}). One can imagine using a network-based framework to purposefully design heterogeneous materials with desired mesoscale topology that gives rise to unique physical properties. To our knowledge, this would be a new perspective, but recent and on-going work in the development of physically informed network tools and models \cite{Bassett2012, Bassett2013Robust, bassett2015extraction, Sarzynska:2015aa} are bringing these objectives closer to fruition.

One approach for material design in a network-based framework may be to utilize multi-objective functions and Pareto optimality \cite{Goldberg:1989}. For instance, starting with a physically realizable material network structure and a known correlation between a network statistic and a material property of interest (mechanical stability, for example), one can rewire the network with a cost function to maximize the network statistic that is correlated with the property of interest and anti-correlated with any properties of non-interest (instabilities, for example). One way to explicitly do the rewiring is along Pareto optimal fronts in a network morphospace \cite{McGhee:1997}, with rewiring rules that preserve network features and topologies required by physical constraints and laws \cite{Avena-Koenigsberger:2014, Avena-Koenigsberger:2013, Goni:2013}, and references therein. This method would theoretically select for physically feasible material network designs while simultaneously optimizing topologies to result in purposefully selected material properties.

\begin{acknowledgements}
The authors gratefully acknowledge support from the John D. and Catherine T. MacArthur Foundation, the Alfred P. Sloan Foundation, the Army Research Laboratory and the Army Research Office through contract numbers W911NF-10-2-0022 and W911NF-14-1-0679,the National Institute of Mental Health (2-R01-DC-009209-11), the National Institute of Child Health and Human Development (1R01HD086888-01), the Office of Naval Research, and the National Science Foundation (BCS-1441502, BCS-1430087, and PHY-1554488).
\end{acknowledgements}

\appendix

\section{Algebraic-topological definition of the TCF}
\label{s:algtop}

While the graph-theoretic definition of the TCF given in Section \ref{s:topology} is sufficient for computation in the case of static 2D packings, to understand the motivation and extend the measurement to the case of dynamic or 3D packings, it is necessary to give a more complete description of the underlying mathematics. These tools are adapted from the field of algebraic topology, which has recently found a broad range of applications; for a more thorough introduction to these methods and their use we recommend \cite{ghrist2014elementary}. 

We begin by describing a canonical combinatorial object one can build from a graph, its \emph{clique complex} (FIG \ref{f:measures}c (left, middle)). Let $G$ be a graph with $N$ vertices, which without loss of generality we take to be the ordered set $[N] = \{1, 2, \dots, N\}$. A \emph{clique} in a graph is a complete (all-to-all connected) subgraph, and we denote a clique on vertices $v_0 < v_1 < \cdots < v_k \subset [N]$ uniquely by the string $v_0v_1\dots v_k$. Cliques with $(k+1)$ vertices can be realized as the convex hull of $k$ points, which generically span a $k$-dimensional region, which leads to a choice of index; it is standard to index by dimension rather than number of vertices, and we will follow that convention here. Now, the set of cliques decomposes as a disjoint union of cliques of fixed size, and we define sets
\begin{eqnarray*}
X_k(G) &\defeq& \{\sigma \subset \{1\dots N\} \text{ so that }\\&&\;\;\sigma \text{ is a clique in } G \text{ with (k+1) vertices} \}.
\end{eqnarray*}
For our purposes, the most important of these will be $X_0(G)$, the set of vertices of $G$; $X_1(G)$, its edges; and $X_2(G)$, its triangles, as no larger cliques can appear in a planar network. However, the underlying principle can be generalized to any network, and these may possess larger cliques. 
 
Now, construct a sequence of vector spaces and linear maps \footnote{These definitions can be made over any field. However, in the restrictive case of a planar network, the choice of field does not affect results of the computation. For computation purposes we will use the field with two elements, allowing us to ignore signs and use faster software implementations.} called the \emph{clique chain complex} of $G$,
\begin{equation*}
C_\bullet^X(G)\; \defeq\; C_0^X(G) \labelleftarrow{$\delta_1$} C_1^X(G) \labelleftarrow{$\delta_2$} C_2^X(G) \labelleftarrow{$\delta_3$} C_3^X(G) \cdots.
\end{equation*}
Define the vector spaces $C_k^X(G)$ to be a vector space with basis elements $\{e_{v_0v_1\dots v_i} \text{ s.t. } v_0v_1\dots v_i \in X_i(G)\}$ corresponding to cliques in $G$. Observe that removing any vertex (and all attached edges) from a clique results in a smaller clique, and that geometrically the collection of all sub-cliques on one fewer vertex forms the boundary of the original clique. This geometry provides us with the definition of the {\em boundary} maps $\delta_k:C_k^X(G) \to C_{k-1}^X(G)$, defined on basis vectors as an alternating sum\footnote{The choice of ordering on the vertices induces a canonical notion of orientation on cliques, and the boundary elements alternate in their orientations. This distinction disappears when working over the field with two elements.} of all sub-cliques with one vertex fewer, 
\begin{equation*}
\delta_i(e_{v_0v_2\dots v_i}) \defeq \sum_{k=0}^i (-1)^ke_{v_0v_1\dots \widehat{v}_{k} \dots v_i},
\end{equation*}
where the hat denotes omission of a vertex, and extended linearly to all of $C_i^X(G)$. By convention, define $\delta_0$ to be the zero map.

There are two distinguished classes of elements in the vector spaces in the chain complex: \emph{cycles}, which are elements in the kernel (or null space) of some $\delta$ map, and \emph{boundaries}, which are elements in the image of some $\delta$ map. It turns out that all boundaries are cycles: this is perhaps clear geometrically, but one can simply compute that, $\delta_{k-1} \circ \delta_{k}$ is the zero map for each $k$:
\begin{equation*}
\begin{split}
  \delta_{k-1}\circ \delta_{k}(e_{v_0v_1\dots v_k})\hspace{0.3\textwidth} \\ = \sum_{i\neq \ell} (-1)^i(e_{v_0v_1\dots \widehat{v}_i \dots \widehat{v}_\ell \dots v_k} - e_{v_0v_1\dots \widehat{v}_i \dots \widehat{v}_\ell \dots v_k})\\  =0\hspace{0.35\textwidth},
\end{split}
\end{equation*}
Thus, the image of $\delta_{k+1}$ is a vector subspace of the kernel of $\delta_{k}$, and we define the $k$th homology group of the clique chain complex to be the quotient vector space
\begin{equation*}
H_k^X(G) \defeq \frac{\text{ker}\delta_k}{\text{im}\delta_{k+1}}.
\end{equation*}
Elements of the $k$th homology group are equivalence classes of $k$-cycles, and the members of each equivalence class differ by boundaries of $(k+1)$-cliques. In the two-dimensional setting, there is a canonically defined minimal representative of each non-zero equivalence class, which we take to correspond to a fundamental circuit in the force chain (FIG. \ref{f:measures}c (right, yellow)). Compact regions will induce cycles that consist of cliques, and these will be equivalent to the trivial cycle (FIG. \ref{f:measures}c (right, blue)).

While this machinery in general provides a useful framework in which to discuss force chain topology, we note that the clique homology groups as defined do not fully capture the interesting branching structure of the graph. In particular, branches that do not close to form circuits are ignored (FIG. \ref{f:measures}c (right, red)). To repair this defect, we consider instead the quotient of the clique complex obtained by identifying all of the leaves $\{\ell_1, \ell_2, \dots \ell_k\} \subseteq [N]$ (FIG. \ref{f:measures}c (middle, green)) to a single point, which creates new cycles that correspond to such branches (FIG. \ref{f:measures}c (right, purple). However, the resulting object may not be a clique complex (if this would result in paths between leaves of length two or less, there would be multiple edges between adjacent vertices); to avoid this complexity, we can instead take the quotient on the level of the chain complex, defining a new chain complex
\begin{equation*}
C_\bullet^{X, \ell}(G)\; \defeq\; C_0^{X, \ell}(G) \labelleftarrow{$\delta_1^\ell$} C_1^X(G) \labelleftarrow{$\delta_2$} C_2^X(G) \labelleftarrow{$\delta_3$} C_3^X(G) \cdots.
\end{equation*}
where $C_0^{X, \ell}(G) = C_0^X(G) / \langle e_{\ell_1} - e_{\ell_2}, e_{\ell_1} - e_{\ell_2}, \dots e_{\ell_1} - e_{\ell_k} \rangle$ is the quotient vector space in which all basis elements corresponding to leaves are identified to a single basis vector and $\delta_1^\ell$ is the map induced on the quotient from $\delta_1$. Now, define the \emph{leaf-reduced clique homology groups}, $H_k^{X,\ell}(G)$ to be the homology groups of this chain complex.

Finally, we consider a numerical measurement of these loops, the \emph{leaf-reduced clique Betti numbers} of $G$, given by
\begin{equation*}
\beta_k^{X,\ell}(G) \defeq \text{rank}(H_k^{X, \ell}(G)).
\end{equation*}
Because the potential size of a homology group scales non-linearly with the size of the underlying graph, it is expedient to look at an appropriate normalization. The presence of cliques corresponds to the presence of compact regions, and thus makes force chains more stable, while the branch points corresponding to non-trivial cycles reduce stability, so we define the \emph{topological compactness factor} to be
\begin{equation*}
TCF(G) \defeq 1 - \frac{\beta_1^{X,\ell}(G)}{\text{rank}(\text{ker}(\delta_{1}^\ell))}
\end{equation*}
which corresponds to the graph-theoretic description given in Section \ref{s:topology}. 

There are several advantages to this linear-algebraic approach. First, it is straightforward to define an analogous statistic for the second (and higher) Betti numbers for use in the 3D case or for use in non-spatial networks. More interesting, however, is the fact that graph homomorphisms induce linear maps on homology. These maps, in turn, can provide information about the evolution of the topological structure of the system under, for example, shear stress or chagnge in packing density.

\end{document}